\documentclass[12pt]{article}
\usepackage{a4p}
\usepackage{epsfig}
\usepackage{amssymb}
\usepackage{amsmath}
\usepackage{cite}
\def\opalackerstaff{OPAL Collaboration, K.\ Ackerstaff \etal}
\def\opalabbiendi{OPAL Collaboration, G.\ Abbiendi \etal}

\def\opalacton{OPAL Collaboration, P.D.\ Acton \etal}

\def\opalahmet{OPAL Collaboration, K.\ Ahmet \etal}
\newcommand{\PLB}[3]  {Phys.\ Lett.\ \textbf{B#1} (#2) #3}
\newcommand{\ZPC}[3]  {Z.\ Phys.\ \textbf{C#1} (#2) #3}
\newcommand{\EPC}[3]  {Eur.\ Phys.\ J.\ \textbf{C#1} (#2) #3}
\newcommand{\NIMA}[3] {Nucl.\ Instr.\ and Meth.\ \textbf{A#1} (#2) #3}

\newcommand{\PRL}[3]  {Phys.\ Rev.\ Lett.\ \textbf{#1} (#2) #3}
\newcommand{\PRD}[3]  {Phys.\ Rev.\ \textbf{D#1} (#2) #3}
\newcommand{\NPB}[3]  {Nucl.\ Phys.\ \textbf{B#1} (#2) #3}

\newcommand{\CPC}[3]  {Comp.\ Phys.\ Comm.\ \textbf{#1} (#2) #3}
\newcommand{\Nobs}{\mbox{$187$}}
\newcommand{\NexpKY}{\mbox{$188.4\pm1.0$}}

\newcommand{\epem}{\mbox{$\mathrm{e^+ e^-}$}}

\newcommand{\mpmm}{\mbox{$\mu^+\mu^-$}}

\newcommand{\tptm}{\mbox{$\tau^+\tau^-$}}

\newcommand{\Zzero}{\mbox{${\mathrm{Z}^0}$}}
\newcommand{\Zgamma}{\mbox{${\mathrm{Z}^0} / \gamma$}}
\newcommand{\WW} {\mbox{$\mathrm{W^+W^-}$}}

\newcommand{\Wmv} {\mbox{$W^{\mu\nu}$}}

\newcommand{\Wma} {\mbox{$W_{\mu\alpha}$}}
\newcommand{\Wka} {\mbox{$W^{(k)\alpha}$}}
\newcommand{\Fmv} {\mbox{$\mathrm{F}^{\mu\nu}$}} 
\newcommand{\FMV} {\mbox{$\mathrm{F}_{\mu\nu}$}} 
\newcommand{\Fma} {\mbox{$\mathrm{F}^{\mu\alpha}$}} 
\newcommand{\FMB} {\mbox{$\mathrm{F}_{\mu\beta}$}} 

\newcommand{\WM} {\mbox{     $\vec{{W}}_{\mu}$  }}
\newcommand{\Wa} {\mbox{     $\vec{{W}}^\alpha$ }}
\newcommand{\WA} {\mbox{     $\vec{{W}}_\alpha$ }}

\newcommand{\Wb} {\mbox{     $\vec{{W}}^\beta$  }}

\newcommand{\WV} {\mbox{$W_{\nu}$}}

\newcommand{\WWg}{\mbox{\WW$\gamma$}}
\newcommand{\sigWWg}{\mbox{$\hat{\sigma}_{\mathrm{WW}\gamma}$}}

\newcommand{\effWWg}{\mbox{$\varepsilon_{\mathrm{WW}\gamma}$}}
\newcommand{\cWWg}{\mbox{$c_{\mathrm{WW}\gamma}$}}

\newcommand{\sigback}{\mbox{$\sigma_{\mathrm{BGD}}$}}

\newcommand{\ZZ}{\mbox{\Zzero\Zzero}}

\newcommand{\qq}{\mbox{$\mathrm{q\overline{q}}$}}

\newcommand{\lpnu}{\mbox{$\ell^+{\nu}_{\ell}$}}
\newcommand{\lmnu}{\mbox{$\ell^-\overline{\nu}_{\ell}$}}
\newcommand{\lnu}{\mbox{$\ell\overline{\nu}_{\ell}$}}

\newcommand{\enu}{\mbox{$\mathrm{e\overline{\nu}_{e}}$}}

\newcommand{\WWqqqq}{\mbox{\WW$\rightarrow$\qq\qq}}
\newcommand{\WWgqqqq}{\mbox{\WW$(\gamma)\rightarrow$\qq\qq$\gamma$}}

\newcommand{\qqqq}{\mbox{\qq\qq}}

\newcommand{\qqlv}{\mbox{\qq\lnu}}
\newcommand{\qqln}{\mbox{\qq\lnu}}
\newcommand{\lnln}{\mbox{\lpnu\lmnu}}
\newcommand{\WWqqln}{\mbox{\WW$\rightarrow$\qq\lnu}}
\newcommand{\WWgqqln}{\mbox{\WW$(\gamma)\rightarrow$\qq\lnu$\gamma$}}

\newcommand{\Wev}{\mbox{$\mathrm{W}$\enu}}

\newcommand{\WWlnln}{\mbox{\WW$\rightarrow$\lpnu\lmnu}}

\newcommand{\WWglnln}{\mbox{\WW$(\gamma)\rightarrow$\lpnu\lmnu$\gamma$}}

\newcommand{\Mzs}{\mbox{$M^2_{\mathrm{Z}^0}$}}
\newcommand{\Mz}{\mbox{$M_{\mathrm{Z}^0}$}}
\newcommand{\Mw}{\mbox{$M_{\mathrm{W}}$}}

\newcommand{\Mws}{\mbox{$M^2_{\mathrm{W}}$}}
\newcommand{\Mfafb}{\mbox{$M_{f_1\overline{f}_2}$}}
\newcommand{\Mfcfd}{\mbox{$M_{f_3\overline{f}_4}$}}
\newcommand{\fafb}{\mbox{$f_1\overline{f}_2$}}
\newcommand{\fcfd}{\mbox{$f_3\overline{f}_4$}}
\newcommand{\Gw}{\mbox{$\Gamma_{\mathrm{W}}$}}

\newcommand{\Egam}{\mbox{$E_\gamma$}}

\newcommand{\Opal}{\mbox{OPAL}}

\newcommand{\cosg}{\mbox{$\cos\theta_\gamma$}}
\newcommand{\cosgl}{\mbox{$\cos\theta_{\gamma \ell}$}}

\newcommand{\cosl}{\mbox{$\cos\theta_{\ell}$}}
\newcommand{\cosgf}{\mbox{$\cos\theta_{\gamma f}$}}

\newcommand{\cosgj}{\mbox{$\cos\theta_{\gamma-\mathrm{JET}}$}}
\newcommand{\GeV}{\mbox{$\mathrm{GeV}$}}

\newcommand{\roots}{\mbox{$\sqrt{s}$}}
\newcommand{\rootsprime}{\mbox{$\sqrt{s'}$}}

\newcommand{\KoralW}{\mbox{KORALW}}
\newcommand{\KandY}{\mbox{KandY}}

\newcommand{\YFSWW}{\mbox{YFSWW}}
\newcommand{\YFSWWT}{\mbox{YFSWW3}}

\newcommand{\KKFF}{\mbox{KK2F}}
\newcommand{\KKff}{\mbox{KK2F}}

\newcommand{\Pythia}{\mbox{PYTHIA}}

\newcommand{\JETSET}{\mbox{JETSET}}

\newcommand{\eewwgammamc}{\mbox{EEWWG}}
\newcommand{\eewwg}{\mbox{EEWWG}}

\newcommand{\RacoonWW}{\mbox{RacoonWW}}

\def\etal{\mbox{{\it et al.}}}
\begin{document}

\begin{titlepage}

\begin{center}
{\Large EUROPEAN ORGANIZATION FOR NUCLEAR RESEARCH}
\end{center}

\begin{flushright}
\large
CERN-EP/2003-043 \\
\ 11 July 2003
\end{flushright}
\bigskip
\begin{center}
 \huge{\bf \boldmath A Study of \WWg\ Events at LEP}
\end{center}
\bigskip
\begin{center}
{ \LARGE The OPAL Collaboration}
\end{center}
\bigskip

\begin{center}{\large Abstract}\end{center}
A study of \WW\ events accompanied by hard photon radiation, 
$E_\gamma>2.5$~GeV,
produced in $\epem$ collisions at LEP is presented. Events consistent
with being two on-shell W-bosons and an isolated photon are selected from
681~pb$^{-1}$ of data recorded at $180~\GeV<\roots<209$~GeV. 
From the sample of \Nobs\ selected \WWg\ candidates 
with
photon energies greater than 2.5~GeV, the 
$\WW\gamma$ cross-section is determined at five values of $\roots$. 
The results are consistent with the Standard Model expectation. 
Averaging over all energies, the ratio of the observed cross-section
to the Standard Model expectation is
\begin{eqnarray*}
    R({\mathrm{data/SM}})    & = & 0.99 \pm 0.09 \pm 0.04,
\end{eqnarray*}
where the errors represent the statistical and systematic uncertainties
respectively. 
These data provide constraints on the related 
${\cal{O}}(\alpha)$ systematic uncertainties on the 
measurement of the W-boson mass at LEP. Finally, the data are used to derive 
95~\% confidence level upper limits 
on possible anomalous contributions to the
$\WW\gamma\gamma$ and $\WW\Zzero\gamma$ vertices:
\begin{eqnarray*}
   -0.020~\mathrm{GeV}^{-2} < & a_0/ \Lambda^2 & < 0.020~\mathrm{GeV}^{-2}, \\ 
   -0.053~\mathrm{GeV}^{-2} < & a_c/ \Lambda^2 & < 0.037~\mathrm{GeV}^{-2}, \\
   -0.16~\mathrm{GeV}^{-2} < & a_n/ \Lambda^2 & < 0.15~\mathrm{GeV}^{-2},
\end{eqnarray*}
where $\Lambda$ represents the energy scale for new physics and $a_0$,
$a_c$ and $a_n$ are dimensionless coupling constants.

\begin{center}
  {\large (To be Submitted to Physics Letters B)}
\end{center}
\end{titlepage} 

\begin{center}{\Large        The OPAL Collaboration
}\end{center}\bigskip
\begin{center}{
G.\thinspace Abbiendi$^{  2}$,
C.\thinspace Ainsley$^{  5}$,
P.F.\thinspace {\AA}kesson$^{  3}$,
G.\thinspace Alexander$^{ 22}$,
J.\thinspace Allison$^{ 16}$,
P.\thinspace Amaral$^{  9}$, 
G.\thinspace Anagnostou$^{  1}$,
K.J.\thinspace Anderson$^{  9}$,
S.\thinspace Arcelli$^{  2}$,
S.\thinspace Asai$^{ 23}$
% ...
}\end{center}\bigskip
\bigskip
$^{  1}$School of Physics and Astronomy, University of Birmingham,
Birmingham B15 2TT, UK
\newline
$^{  2}$Dipartimento di Fisica dell' Universit\`a di Bologna and INFN,
I-40126 Bologna, Italy
\newline
$^{  3}$Physikalisches Institut, Universit\"at Bonn,
D-53115 Bonn, Germany
\newline
% ...
$^{  5}$Cavendish Laboratory, Cambridge CB3 0HE, UK
\newline
% ...
$^{  9}$Enrico Fermi Institute and Department of Physics,
University of Chicago, Chicago IL 60637, USA
\newline
% ...
$^{ 16}$Department of Physics, Schuster Laboratory, The University,
Manchester M13 9PL, UK
\newline
% ...
$^{ 22}$Department of Physics and Astronomy, Tel Aviv University,
Tel Aviv 69978, Israel
\newline
$^{ 23}$International Centre for Elementary Particle Physics and
Department of Physics, University of Tokyo, Tokyo 113-0033, and
Kobe University, Kobe 657-8501, Japan
\newline
% ...
$^{  *}$ Deceased
\bigskip

\section{Introduction}

The \WW\ pair production 
cross-section has been precisely measured at LEP over a range of
centre-of-mass energies~\cite{bib:xs189,bib:adlxs,bib:lepxs}. 
The data are well described by the Standard Model (SM) 
expectation~\cite{bib:YFSWW,bib:RacoonWW}. The good agreement between
experiment and theory is only obtained once factorizable and
non-factorizable $\cal{O}(\alpha)$ photonic corrections are 
included in the theoretical calculations (see for example \cite{bib:LEP4F} and
references therein). The inclusion
of real and virtual photonic corrections in the \YFSWW\cite{bib:YFSWW}
and \RacoonWW\cite{bib:RacoonWW} programs has 
reduced the theoretical uncertainty on the CC03 $\epem\rightarrow\WW$
cross-section\footnote{CC03 refers to the
three doubly resonant diagrams for $\epem\rightarrow\WW$.} to below 0.5~\%~\cite{bib:LEP4F}. 
Uncertainties in these $\cal{O}(\alpha)$ corrections may 
lead to small, but non-negligible, systematic uncertainties in the 
determination of the W-boson mass, $\Mw$, at LEP\cite{bib:sysMw}. 
This paper presents a study of the process 
$\epem\rightarrow\WW\gamma$ and thus probes the modelling of 
real photonic corrections to the \WW\ pair creation process. 
The data are used to obtain measurements of the $\epem\rightarrow\WW\gamma$
cross-section within a restricted phase-space region, $\sigWWg$,  
for $180~\GeV<\roots<209$~GeV. 

In the SM, photon radiation in the $\WW$ production process 
at LEP can be categorized into four main classes of diagrams: 
initial state radiation (ISR); final state radiation (FSR) from a lepton; 
FSR from the quark or from the associated parton shower; 
and bremsstrahlung from one of the intermediate W-bosons, referred to as 
WSR. At LEP energies WSR has a 
significant effect only through interference with ISR.  
Experimentally photons arising from decays of hadrons in a jet are
indistinguishable from FSR photons from a quark or parton shower. 
For this reason, and due to the relatively large uncertainties
in the Monte Carlo modelling of photon production in the parton shower, 
all photons associated with hadronic jets (from hadron decay and FSR) are 
considered background for the measurements of $\sigWWg$. 
 
The measurements of the $\WWg$ cross-section are compared with the 
predictions of the \KoralW~\cite{bib:KoralW}, \KandY~\cite{bib:KandY} 
(the concurrent Monte Carlo \KoralW1.51\ and \YFSWWT) 
and \RacoonWW~\cite{bib:RacoonWW} programs.  These comparisons are used 
to obtain the first data-driven estimate of the 
systematic uncertainty on $\Mw$ due to the Monte Carlo description of 
real photon radiation in $\WW$ events. 

In addition, the $\WWg$ final state is sensitive to possible
anomalous $\WW\gamma\gamma$ and $\WW\Zzero\gamma$
quartic gauge boson couplings (QGCs). At LEP energies the contribution of
the SM QGC diagram is negligible. The data presented in this paper
are used to place upper limits on the size of possible anomalous QGCs. 
These limits are more than a factor three tighter than previous
\Opal\ results from $\epem\rightarrow\WWg$\cite{bib:OPALwwg} and
are consistent with other measurements\cite{bib:L3wwg}.

\section{The OPAL Detector, Data Samples and Monte Carlo}

\subsection{The OPAL Detector}

The \Opal\ detector includes a 3.7 m diameter tracking volume within
a 0.435~T axial magnetic field. The tracking detectors include a silicon
micro-vertex detector, a high precision gas vertex detector and a large
volume gas jet chamber. The tracking acceptance corresponds to approximately
$|\cos\theta|<0.95$ (for the track quality cuts used in this study)\footnote{The \Opal\ right-handed
                         coordinate system is defined such that the
                         origin is at the centre of the detector and the
                         $z$ axis points along the direction of the $e^-$
                         beam; $\theta$ is the polar angle with respect to 
                         the $z$ axis.}.
Lying outside the solenoid, 
the electromagnetic calorimeter (ECAL) 
consisting of $11\,704$ lead glass blocks
has  full acceptance in the range $|\cos\theta|<0.98$ and a relative
energy resolution of approximately $6~\%$ for 10~GeV photons.
The magnet return yoke is instrumented with streamer tubes which
serve as the hadronic calorimeter. Muon chambers outside the 
hadronic calorimeter provide muon identification in the range
$|\cos\theta|<0.98$.  A detailed description of the \Opal\ detector can be
found in \cite{bib:detector}.  

\subsection{Data Sample}

During LEP\,2 operation the centre-of-mass energy was increased from 
161~GeV to 209~GeV in several steps. The total integrated luminosity of 
the data sample considered in this paper, evaluated using small angle 
Bhabha scattering events observed in the silicon tungsten forward 
calorimeter~\cite{bib:lumi}, is $(681\pm2)~\mathrm{pb}^{-1}$.
For the purpose of measuring the $\WWg$ cross-section these data
are divided into the five  $\roots$ ranges listed in 
Table~\ref{tab:lumi}. These ranges reflect the main energy steps as
the centre-of-mass energy was increased during LEP2 operation. 
The data recorded at 
161~GeV and 172~GeV, corresponding to a total integrated luminosity
of  20~pb$^{-1}$, are not used here.

\begin{table}[htbp]
\renewcommand{\arraystretch}{1}
\begin{center}
\begin{tabular}{|c|c|c|} \hline 
     Range/GeV  & $<\roots>$/GeV  & $\cal{L}$/pb$^{-1}$ \\ \hline
   180.0$-$185.0  &   182.68    &  57.2     \\
   188.0$-$189.0  &   188.63    & 183.1     \\
   191.0$-$196.0  &   194.44    & 105.7     \\
   199.0$-$204.0  &   200.21    & 114.1     \\
   204.0$-$209.0  &   205.92    & 220.6     \\ \hline
\end{tabular}
\end{center}
\caption{The energy binning used for the \WWg\ cross-section measurements.
         The $\roots$ range covered by each bin, the mean luminosity weighted 
         value of $\roots$ and the corresponding integrated luminosity, 
         ${\cal{L}}$, are listed. 
\label{tab:lumi} }
\end{table}

\subsection{Monte Carlo}

A number of Monte Carlo (MC) samples, all including a 
full simulation\cite{bib:GOPAL} of the \Opal\ detector, 
are used to simulate the SM signal and background processes. 
For this paper the main MC samples for the
process $\epem\rightarrow\WWg$ were generated using the 
\KandY~\cite{bib:KandY} program and, 
unless otherwise specified, the SM expectations for the
$\epem\rightarrow\WWg$ cross-section refer to the \KandY\ prediction.
\KandY\ includes exact $\cal{O}(\alpha)$ YFS 
exponentiation\cite{bib:YFS} for the \WW\ production process, 
with $\cal{O}(\alpha)$ electroweak non-leading (NL) corrections combined 
with YFS exponentiated ${\cal{O}}(\alpha^3)$ leading logarithm (LL) 
initial state radiation. Final state radiation from leptons 
is implemented in PHOTOS~\cite{bib:Photos}  and radiation from
the quark induced parton-shower is performed by JETSET~\cite{bib:Jetset}. 
The most notable improvements over the \KoralW\ program are the 
leading non-factorizable corrections in the Screened Coulomb 
ansatz\cite{bib:Chapovsky}, the inclusion of bremsstrahlung from the
W-pairs (WSR), and the implementation of
$\cal{O}(\alpha)$ electroweak NL corrections.

The \KoralW\ program\cite{bib:KoralW} is used to simulate the background
from four-fermion final states which are incompatible with
coming from the decays of two W-bosons ({\em e.g.} $\epem\rightarrow\qq\mpmm$). 
The two-fermion background processes,
 $\epem\rightarrow\Zzero/\gamma\rightarrow\qq$ 
and $\epem\rightarrow\Zzero/\gamma\rightarrow\tptm$, are 
simulated using \KKFF~\cite{bib:kk}. The background in the
$\WWg$ event selection from multi-peripheral two-photon diagrams
was found to be negligible.   

In addition, the $\RacoonWW$ program\cite{bib:RacoonWW} is 
used in the Improved Born Approximation (IBA) mode to obtain 
independent predictions of the cross-sections for 
$\epem\rightarrow\WW\gamma$ and
$\epem\rightarrow4f\gamma$. In this mode all lowest order diagrams
contributing to $\epem\rightarrow\WW\gamma$ are included. 
The $\eewwgammamc$ program\cite{bib:eewwg} is used to 
obtain predicted cross-sections in the presence of anomalous
QGCs which are then used to extract experimental limits on the 
anomalous contributions to the $\WW\gamma\gamma$ and $\WW\Zzero\gamma$ 
vertices.

\section{\boldmath $\WW\gamma$ Signal Definition}
\label{sec:sigdef}

The process  $\epem\rightarrow\WW\gamma$ results in a four-fermion plus 
photon final state, $\fafb\fcfd\gamma$, where the fermion flavours are
appropriate for W-decay. In the SM, photons are radiated in several
classes of diagrams corresponding to ISR, 
FSR from both charged leptons and quarks, 
radiation from the W-boson (WSR) and the Standard Model QGC diagram. 
The invariant mass distributions of the fermions
are different for the different radiation processes. In the case of ISR,
the $\fafb$ and $\fcfd$ systems are produced with invariant
masses close to \Mw. In the case of FSR,
the $\fafb\gamma$ and $\fcfd$ combinations or the $\fafb$ and 
$\fcfd\gamma$ combinations give invariant masses close to the 
W-boson mass. For photon energies $\Egam>\Gw$, where $\Gw$ is 
the W-boson width, events from FSR tend to occupy a different
kinematic region from those arising from the ISR or QGC diagrams.
Consequently, interference between FSR and ISR/QGC diagrams 
is suppressed.  At LEP energies the effect of WSR is only significant
through interference with ISR; the WSR diagrams are
only of relevance to the region of phase-space populated by ISR diagrams.   

Only part of the $\WWg$ phase-space is accessible experimentally and, 
therefore,  it is necessary to define
a specific region of phase-space in which the cross-section will be measured.
The definition of the signal region is chosen to be well matched to the 
experimental sensitivity. In addition, by defining the cross-section to 
correspond to a region of four-fermion phase-space dominated by the 
doubly resonant $\WW$ production (CC03) diagrams, 
contributions from other interfering diagrams
can be made small. In this way, the experimental results can be 
compared with both the predictions of calculations implementing all
four-fermion diagrams and with calculations implementing only
CC03 diagrams. Finally, 
invariant mass cuts are imposed to reduce the contribution of 
FSR both from quarks and from leptons. This is desirable for two 
reasons. Firstly, any new physics is unlikely to manifest itself in a 
modification of FSR. Secondly, it reduces modelling uncertainties which 
are potentially large in the case of FSR from the quark-induced parton 
shower.   

In this paper, the 
$\WW\gamma\rightarrow\fafb\fcfd\gamma$ cross-section, denoted by
$\sigWWg$, is measured for: 
\begin{itemize}
  \item $\Egam   > 2.5$~GeV, where \Egam\ is the photon energy.
  \item $|\cosg| < 0.975$, where $\cosg$ is the cosine of the polar 
                         angle of the photon.
  \item $\cosgf  < 0.90$, where $\cosgf$ is the cosine of the
                         minimum angle between the
                         photon and any of the charged fermions in the
                         four-fermion final state.
  \item $|\cosl|  < 0.95$, where $|\cosl|$ is the modulus of
                         the cosine of the
                         polar angle of the charged lepton in the 
                         $\WW\rightarrow\qqlv$ final state. In the
                         $\WW\rightarrow\lnln$ final state this requirement 
                         applies to both of the charged leptons.
  \item $|\Mfafb-\Mw|$ and $|\Mfcfd-\Mw| < 3\,\Gw$, 
	                where \Mfafb\ and \Mfcfd\ are
                        the invariant masses of fermions consistent with
                        being from the decays of  
                        the W$^-$ or W$^+$. 
\end{itemize}
The first three requirements are closely matched to the ability to
reconstruct a pure sample of isolated photons in the \Opal\ detector.
The requirement on the polar angle of the charged leptons from W-decay
is imposed because the \WW\ event selection becomes significantly less
efficient beyond the acceptance of the tracking chambers. It also reduces
contributions from interfering four-fermion background diagrams such 
as the $t$-channel process $\epem\rightarrow\Wev$.  The cut on the
invariant masses of the fermion pairs further reduces the (interfering) 
four-fermion backgrounds and suppresses the contribution of FSR to 
the signal region. Due to the finite jet width, jets are detected
over the full polar angle acceptance and therefore there is no explicit 
requirement on the polar angle of the quark.

In the above definition of the signal, all requirements are made on
generator level quantities. Generator level refers to the true four-momenta 
of particles in the $\fafb\fcfd\gamma$ final state. 
The cross-section within the above kinematic cuts,
\sigWWg, is dominated by doubly-resonant \WW\ production.
For example, the difference between the cross-section 
for
the full set of $4f\gamma$ diagrams relative to cross-section for the CC03
diagrams alone is less than 0.5~\% (calculated using the 
IBA implemented in \RacoonWW\cite{bib:RacoonWW}).  

\section{\boldmath $\WW\gamma$ Event Selection}

The selection of $\WW\gamma$ events proceeds in three stages:
selection of $\WW$ events, photon identification,
and background rejection using kinematic information. 
All $\WW$ final states are used in this study.

\subsection{\boldmath $\WW$ Selection}

The $\WWlnln$, $\WWqqln$ and $\WWqqqq$ selections of reference \cite{bib:xs189}
are used as the basis of the $\WWg$ selections\footnote{Reference 
\cite{bib:xs189} refers to the event selection at $\roots=189$~GeV.
For data recorded at higher centre-of-mass energies the same likelihood 
selection is used but with reference distributions obtained from Monte 
Carlo events generated at higher centre-of-mass energies.}.
For $\WWglnln$ and $\WWgqqln$ the standard selections are applied.
For $\WWgqqqq$ events, a modified version of
the $\WWqqqq$ selection of reference \cite{bib:xs189} is used. 
In the standard selection, events are forced into four jets using the Durham
$k_{T}$ algorithm~\cite{bib:Durham}. In 
approximately $10~\%$ of Monte Carlo events with high energy photons 
($\Egam>10$~GeV), the photon alone forms one of the
four jets. This introduces an additional inefficiency, due to the
requirement in the preselection that there should be at least
one charged particle track associated with each jet. 
For this reason, events failing the standard
$\WWqqqq$ selection are forced into four jets after excluding the highest
energy isolated electromagnetic calorimeter cluster and the selection
re-applied. 
The overall selection efficiency for $\WW\gamma$ events within the signal
definition is 88~\% and is approximately independent of centre-of-mass
energies for $180~\GeV<\roots<209$~GeV.

\subsection{Photon Identification}

Photon identification is similar to that described in \cite{bib:OPAL_HGG},
although for this study the minimum photon energy is reduced to 2.5~GeV.
Photon candidates are identified as one of three types:
\begin{itemize}
 \item Unassociated ECAL clusters 
       defined by the requirement that no 
       charged particle track, when extrapolated to the front-face of 
       the ECAL, lies within a distance defined by the typical 
       angular resolution of the ECAL cluster. The lateral 
       spread of the 
       cluster was required to satisfy the criteria described in 
       reference \cite{bib:OPAL_HGG}. 
 \item Two-track photon conversions which 
       are selected using an artificial 
       neural network as described in \cite{bib:ANN}.
 \item Conversions where only a single track is reconstructed, identified 
       as an electromagnetic calorimeter 
       cluster associated with a track which is consistent 
       with originating from a photon conversion. The track is
       required to have no associated hits 
       in either layer of the silicon micro-vertex detector or in the 
       first six layers of the central vertex chamber.
\end{itemize}
For both types of conversion, the photon energy is defined by the 
sum of cluster energies pointed to by the track(s).

Photon candidates identified using the above criteria are 
required to satisfy isolation requirements. 
The summed energies of any additional tracks and clusters in 
a $20^\circ$ half-angle cone defined by the photon direction 
have to be less than 2 GeV. In addition,
the energy deposited in the hadron calorimeter in a $20^\circ$ half-angle 
cone around the photon candidate is required to be less than 5~GeV.
If the invariant mass formed from 
the photon candidate and the energy deposit in any ECAL cluster is 
less than 0.25~GeV/$c^2$ the candidate is rejected in order to suppress
 photons from $\pi^0$ decay.
For photon candidates with $2.5~\GeV<\Egam<10.0~\GeV$ a relative 
likelihood selection is applied to reduce the background from 
photons from the decays of hadrons (dominated by $\pi^0$ and $\eta$ decays). 
The likelihood is based on five discriminant variables: $\Egam$, $|\cosg|$, 
the angle between the photon and the nearest jet, 
the angle between the photon and the nearest track, 
and the minimum invariant mass formed from the 
photon candidate and any other ECAL cluster in the event.  
For photons above 10~GeV the background is low and no photon
identification likelihood is needed.

\subsection{Photon Acceptance}

The identified photon is required to lie within the polar acceptance,
\begin{itemize}
 \item $|\cosg| < 0.975$.
\end{itemize}
The photon is also required to be isolated from the charged
fermions in the final state.
Cuts are applied on the cosine of the angle between the 
photon and closest jet, $\cosgj$, and on the cosine of the 
angle between the photon and a charged lepton from the W-boson decay, 
$\cosgl$:
\begin{itemize}
 \item $\cosgj < 0.9$ for $\WWqqln\gamma$ and $\WWqqqq\gamma$ events,
 \item $\cosgl < 0.9$ for $\WWlnln\gamma$ and $\WWqqln\gamma$ events.
\end{itemize}

For selected events with photons within the generator level acceptance
the photon identification efficiency is 75~\% for $\Egam \ge 7.5\,\GeV$,
69~\% for $5.0~\GeV \le \Egam<7.5\,\GeV$ and 45~\% for 
$2.5~\GeV \le \Egam < 5.0\,\GeV$. 
The photon identification efficiency is almost independent of $\cosg$
in the region $|\cosg|<0.975$. The non-photonic backgrounds are 
less than 4~\% for $|\cosg|<0.95$. For  $|\cosg|>0.95$ the 
background increases to 8~\%.
If more than one photon candidate passes the photon acceptance requirements
only the highest energy photon is retained for the following analysis. 

\subsection{Kinematic Requirements}
 
The photon in selected $\WWg$ events is classified as 
ISR, FSR from the lepton, or as being associated with a jet
(either FSR from the parton shower or coming from hadron decay).
No special treatment is made for WSR because 
WSR diagrams are only observable through interference with ISR diagrams
and, consequently, the effects of WSR diagrams will be apparent in the 
event sample classified as ISR. 
In $\lnln$ events, photons are classified as ISR if
$\cosgl<|\cosg|$, otherwise the photons are classified as FSR from one 
of the charged leptons. 
For the $\qqlv$ and $\qqqq$ channels
the classification is performed using a relative likelihood selection
in which kinematic fitting plays 
a major r\^{o}le. Three kinematic fits are employed, corresponding
 to the following hypotheses: 
\begin{itemize}
 \item[a)] the photon originates from FSR from the quark; the fit 
           assumes a two-body \WW\ final state, where the identified photon 
           is included as part of the nearest jet.
 \item[b)] the photon originates from FSR from the lepton   
       (only used for \WWgqqln\ events); the fit assumes a two-body 
       \WW\ final state, where the photon is associated with the 
        charged lepton.
 \item[c)] the photon originates from ISR; the fit assumes a 
         three body final state 
         consisting of the two W-bosons and the photon. 
\end{itemize}  
In each case, the constraints of energy and 
momentum conservation are imposed and the two reconstructed 
masses of the W-boson candidates are required to be 
equal~\cite{bib:mass172}. 
An event is considered consistent with one of the above
hypotheses if the fit converges with a fit probability of greater than
0.1~\% and if the reconstructed W-boson mass is greater than 74~GeV.
In fully hadronic events there are three possible jet-pairing 
combinations. Here, for each fit hypothesis, the combination yielding
the highest kinematic fit probability is used. 

\begin{figure}[htbp]
\begin{center}
 \epsfxsize=14cm
 \epsffile{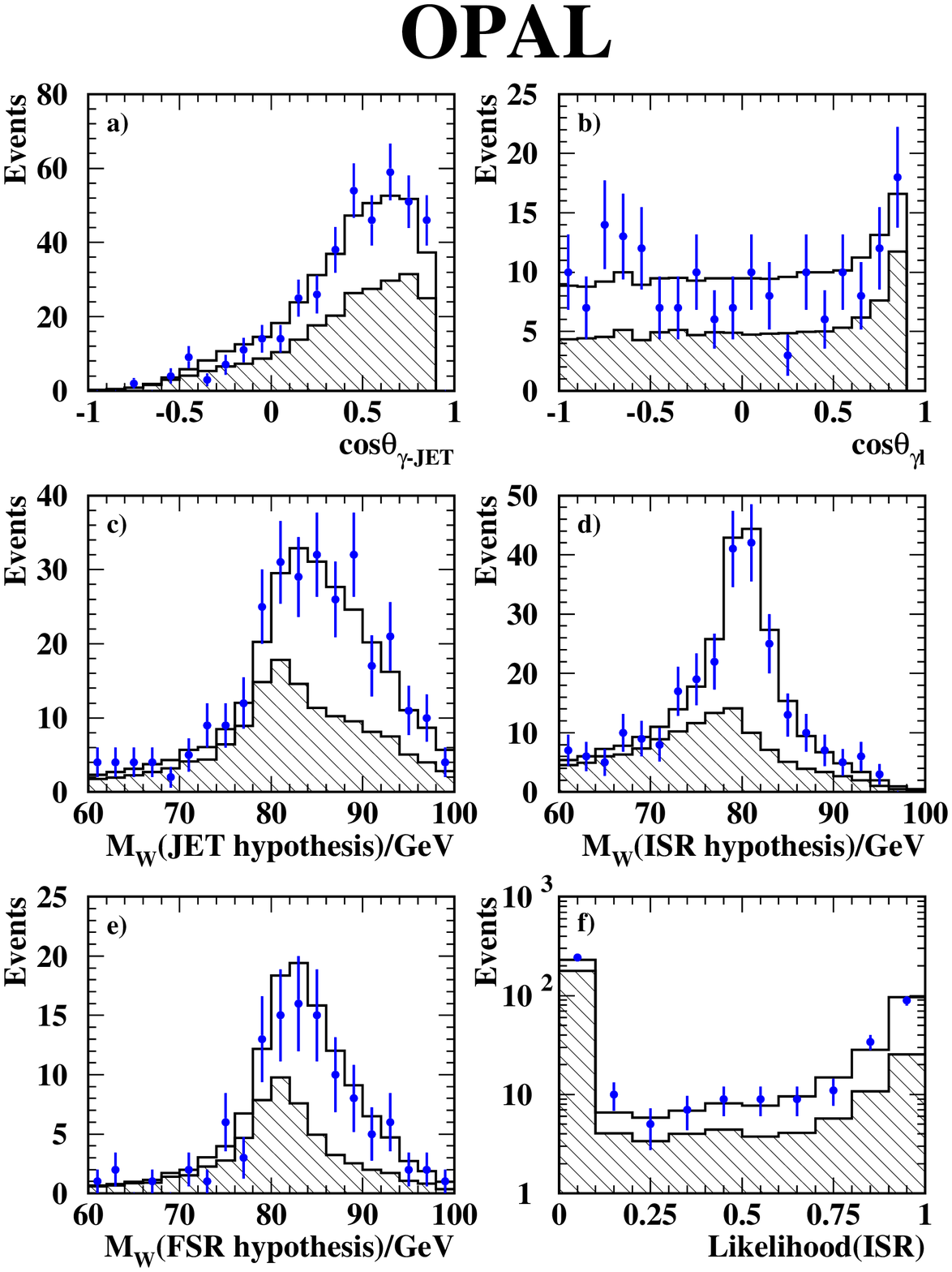}
 \caption{The five kinematic variables used to classify the photon in
          $\WWg$ events as being from ISR, FSR or associated with the jet.
	  Unless otherwise specified the distributions are shown for
          $\qqln$ and $\qqqq$ events combined. 
          The variables are: a) the angle between the photon and the
          nearest jet, \cosgj; b) the angle between the photon and 
          the charged lepton, \cosgl\ (\qqln\ events only); 
              c) the reconstructed
          W-boson mass under the hypothesis that the photon is
          associated with jet; d) the reconstructed
          W-boson mass under the hypothesis that the photon is
          from ISR; e) the reconstructed
          W-boson mass under the hypothesis that the photon is
          from FSR ($\qqln$ only). Plot f) shows the resulting
          relative likelihood distribution for the ISR hypothesis.
          In all cases the data are shown by points with error bars,
          the total SM expectation is shown by the histogram and the
          contributions from processes other than ISR are shown by
          the hatched histograms.     
 \label{fig:like}}
\end{center}
\end{figure}

The reconstructed W-boson mass from the three kinematic fit hypotheses
along with the cosine of the angle between the photon and the nearest
jet are used as the inputs to the relative likelihood.  
For $\qqln$ events the cosine of the angle between the photon and the charged
lepton is also used.  
The distributions used in the relative likelihood classification
are shown in Figure~\ref{fig:like}. Good agreement between data 
and simulation is observed.
Three relative likelihoods are constructed and events are classified as
being either from ISR, FSR from the charged lepton or 
radiation associated with the jets. 
The resulting ISR relative likelihood distribution, ${\cal{L}}_{ISR}$ 
is shown in Figure~\ref{fig:like}f. Events are classified as ISR if
 ${\cal{L}}_{ISR} > {\cal{L}}_{FSR}$ and ${\cal{L}}_{ISR} > {\cal{L}}_{JET}$,
where ${\cal{L}}_{FSR}$ and ${\cal{L}}_{JET}$ are the relative likelihoods
for the respective hypotheses of FSR and radiation associated with the jets.
Only those \WWg\ candidate events classified as ISR 
are retained for the analysis. These events are 
consistent with on-shell W-bosons (fit c),
$\Mfafb\sim\Mfcfd\sim\Mw$ and an isolated photon. 
This procedure suppresses events with final state radiation and events
where the photon is from hadron decay. It also significantly
reduces background from $\epem\rightarrow\qq\gamma$.
As a result the systematic uncertainties from 
photons associated with jets (FSR and $\pi^0$/$\eta$ decays)
are greatly reduced. 
 
The application of the above kinematic requirements 
retains approximately 75~\% of selected signal \WWg\ events with an 
identified photon (using the definition of 
Section \ref{sec:sigdef}) whilst rejecting 
$85~\%-98~\%$ (increasing with the photon energy) of events with photons 
either from FSR or from the decays of mesons.
 
\section{\boldmath Measurement of the \WWg\ cross-section}

\begin{figure}[htbp]
 \epsfxsize=\textwidth
  \epsffile{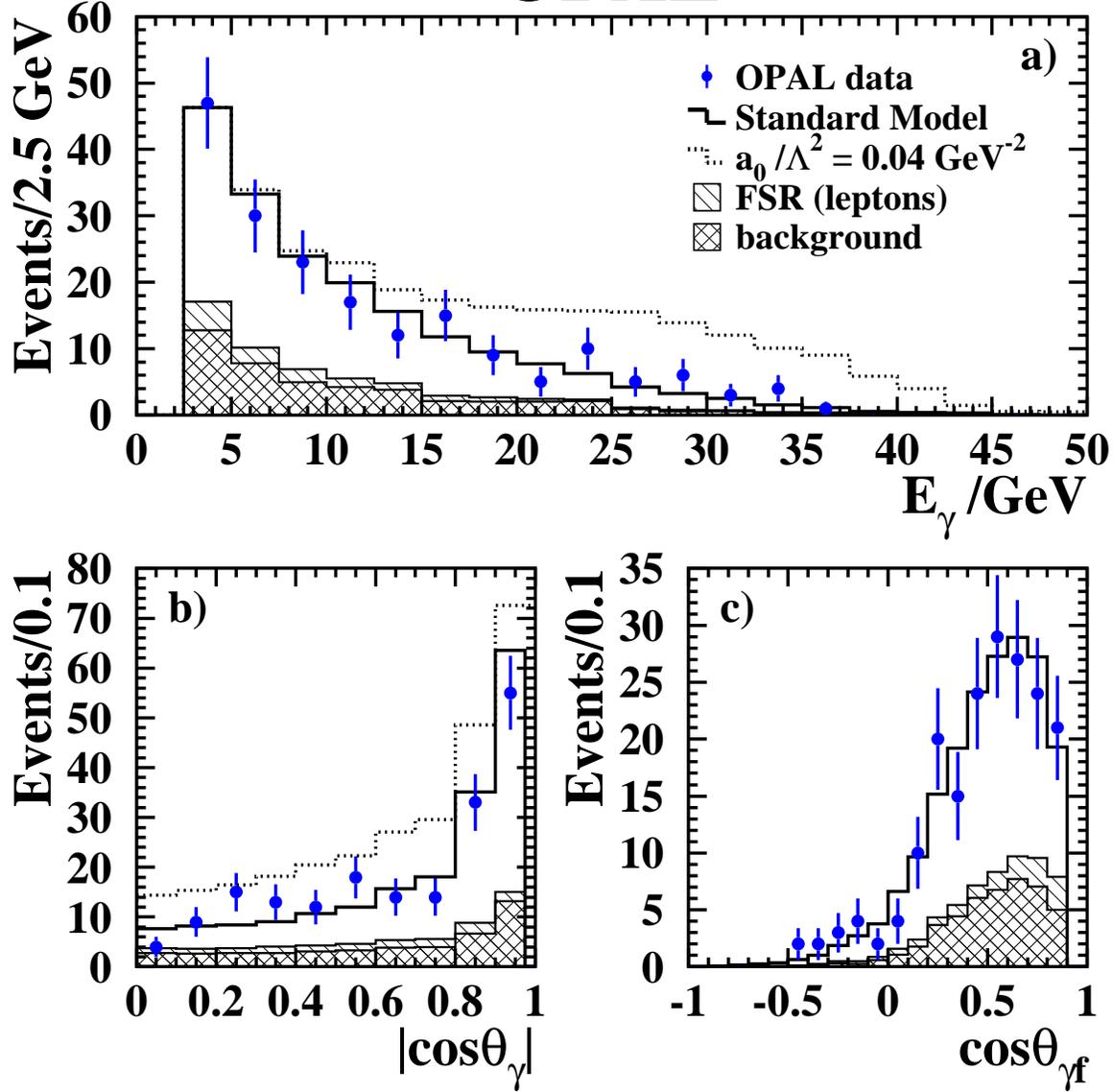}
 \caption{For 
 selected $\WW\gamma$ events $(180~\GeV<\roots<209~\GeV)$,
          a) shows the photon energy spectrum, b) 
          the modulus of the cosine of the polar angle of the
          photon, and c) the cosine of the angle between the photon and 
          the nearest charged fermion. 
          The data are shown by the points
          with error bars and the SM expectations (KandY)
          are shown by the histograms. The doubly-hatched histograms 
          indicate the contributions
          from non-\WW\ background and background from photons
          associated with the parton-shower (either FSR or from hadron 
          decay). The singly-hatched histograms show the contributions 
          from FSR from leptons. The expected $\Egam$  and
          $|\cosg|$ distributions for an anomalous QGC of 
          $a_0/ \Lambda^2 = 0.040~\mathrm{GeV}^{-2}$ are also shown.}
 \label{fig:result_eg}
\end{figure}

Using the selection criteria defined in the previous section,
\Nobs\ $\WWg$ events with $\Egam>2.5$~GeV are selected compared to the 
\KandY\ expectation of \NexpKY\ events (where the error on the expectation
is the quadrature sum of the MC statistical error and luminosity error). 
Figure~\ref{fig:result_eg}a 
shows the photon energy spectrum for the selected $\WW\gamma$ events.
Figure~\ref{fig:result_eg}b shows the distribution
of $|\cosg|$ and Figure~\ref{fig:result_eg}c shows the distribution 
of the cosine of the angle between the photon and the nearest charged 
fermion from the reconstructed W-decay ({\em i.e.} lepton or jet) 
in the event. Good agreement between data and Monte Carlo is 
observed for all distributions. The effect of an
anomalous QGC on the photon energy and polar angle distributions is also
shown.  

\begin{figure}[htbp]
 \epsfxsize=\textwidth
  \epsffile{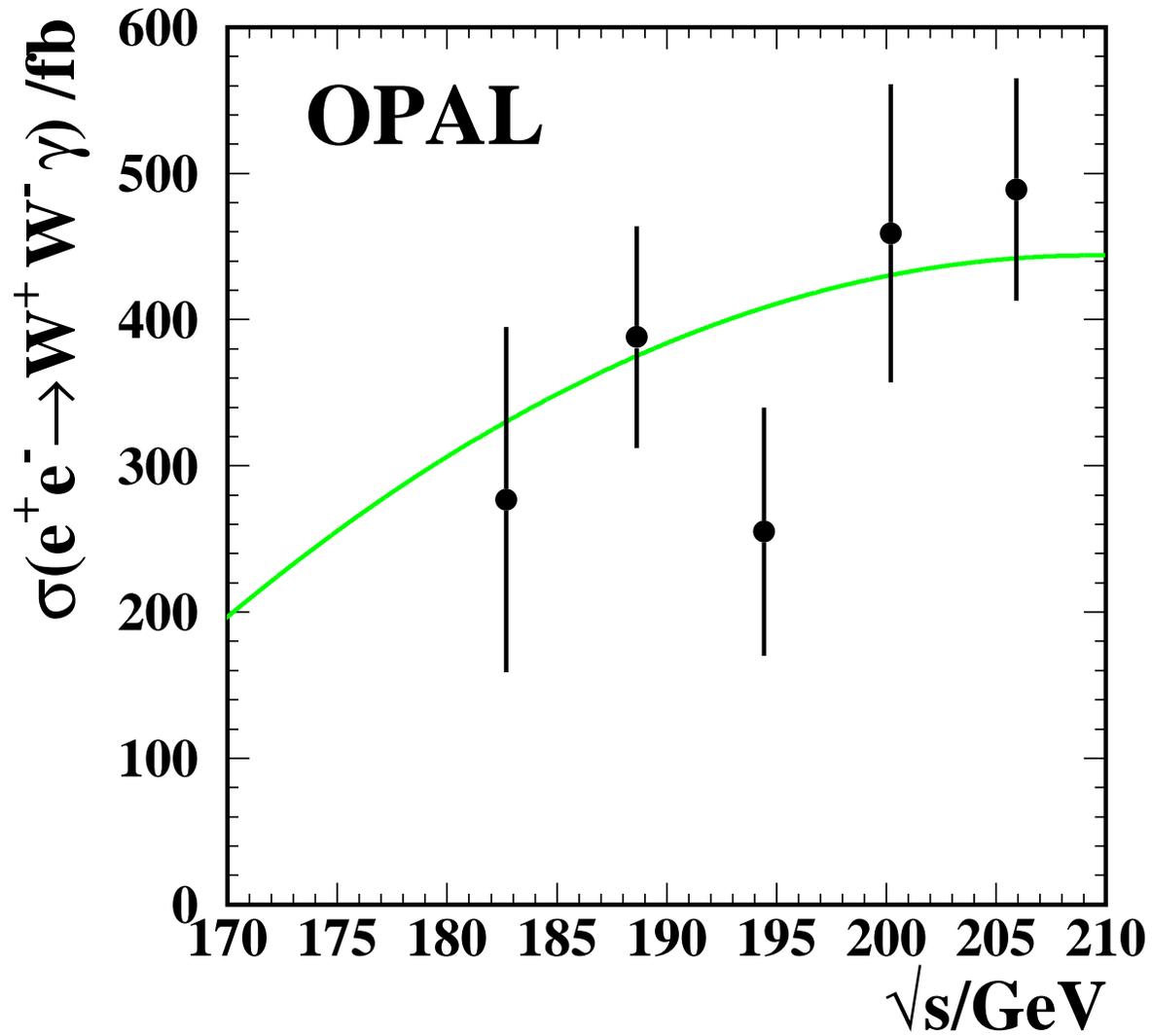}
 \caption{Measured $\WW\gamma$ cross-section for the
          signal definition of Section~\ref{sec:sigdef}.
          The points with error bars show the \Opal\ measurements.
          The curve shows the SM expectation obtained from the
          \KandY\ program.
 \label{fig:xsec}}
\end{figure}

The $\WWg$ cross-section is determined within the acceptance defined
in Section~\ref{sec:sigdef} for the five mean centre-of-mass 
energies listed in Table~\ref{tab:lumi}. 
The $\WWg$ cross-section is calculated from
\begin{eqnarray*}
     \sigWWg & = & \frac{  (N_{\mathrm{obs}}- \sigback{\cal{L}})}
                             { \cWWg\effWWg{\cal{L}}   }  ,
 \label{eqn:sigWWg}
\end{eqnarray*}
where $N_{\mathrm{obs}}$  is the accepted number of events, $\sigback$ is
the SM background cross-section
and ${\cal{L}}$ is the integrated luminosity. The selection efficiency for 
events generated within the acceptance defined in Section~\ref{sec:sigdef}, 
\effWWg, is evaluated using \KandY\ MC $\WWg$ events.
Background from migration of $\WWg$ events from just outside the 
signal region into the selected event sample due to finite 
detector resolution is accounted for by a factor $\cWWg$. This allows
the contribution from selected $\WWg$ events outside the 
signal definition but within the acceptance $\Egam>2.0$~GeV 
and $|\cosg|<0.98$ to scale with the measured cross-section (in contrast
to treating this component as background which is fixed by Monte
Carlo expectation).
The selection efficiency, $\effWWg$, varies from $41~\%-47~\%$ 
increasing with centre-of-mass energy. The correction factor,
$\cWWg$, is 1.14 and is almost independent of centre-of-mass energy.
The background cross-section, \sigback, is estimated using \KandY\ and
\KKff. The background from \WW\ events with photons associated with 
the jets, including photons from FSR from the parton-shower, is scaled
by a factor of $1.30\pm0.15$, as described in Section~\ref{sec:sys}, 
to account for known discrepancies between data and the \JETSET\ prediction. 
The Monte Carlo predicts
that 24~\% of the selected event sample arises from background processes
(including photons associated with jets).
The results are listed in Table~\ref{tab:xsec} where they are 
compared to the predictions from \KandY, and are displayed in 
Figure~\ref{fig:xsec}. The systematic uncertainties
are described in Section~\ref{sec:sys}.  
For the purpose of
combination with the other LEP experiments, the results for a 
more restrictive signal acceptance are given in the Appendix.
\begin{table}[htbp]
\renewcommand{\arraystretch}{1}
\begin{center}
\begin{tabular}{|c|rcrcl|c|} \hline
                   & \multicolumn{6}{c|}{$\sigWWg$/fb} \\ 
   $<\roots>$/GeV  & \multicolumn{5}{c|}{Data} & $\KandY$ \\ \hline
   182.68          & 277 &\hspace{-3mm}$\pm$& \hspace{-4mm}117& \hspace{-3mm}$\pm$& \hspace{-4mm}13 & $327\pm3$  \\
   188.63          & 388 &\hspace{-3mm}$\pm$& \hspace{-4mm} 74& \hspace{-3mm}$\pm$& \hspace{-4mm}17 & $378\pm4$  \\
   194.44          & 255 &\hspace{-3mm}$\pm$& \hspace{-4mm} 84& \hspace{-3mm}$\pm$& \hspace{-4mm}15 & $411\pm4$  \\
   200.21          & 459 &\hspace{-3mm}$\pm$& \hspace{-4mm}100& \hspace{-3mm}$\pm$& \hspace{-4mm}20 & $427\pm4$  \\
   205.92          & 489 &\hspace{-3mm}$\pm$& \hspace{-4mm} 73& \hspace{-3mm}$\pm$& \hspace{-4mm}21 & $443\pm4$  \\ \hline
\end{tabular}
\end{center}
\caption{\WWg\ cross-section measurements for the five centre-of-mass
         energies listed in Table \ref{tab:lumi}. The errors on the 
         measurements are statistical and systematic respectively.
         The errors on the KandY expectations are due to limited
         Monte Carlo statistics.
\label{tab:xsec} }
\end{table}

Table~\ref{tab:ratio} shows the ratio of measured to predicted $\WWg$
cross-sections averaged over the five values of $\roots$ for the
theoretical predictions from \KandY, \RacoonWW, \eewwg\ and \KoralW. 
For \RacoonWW, FSR from the parton shower is included as signal since,
unlike for \KandY\ and \KoralW, there is no way of removing its 
contribution at the generator level. As a consequence of the uncertainties
of the modelling of photons from the parton shower this results in
an increased systematic uncertainty as discussed in Section~\ref{sec:sys}.  
For \eewwg, which does not include any FSR, the expectation for
the contribution from FSR from leptons (which is considered signal)
is taken from PHOTOS. The experimental results correspond to a
measurement with $10~\%$ precision 
of the $\WWg$ cross-section. The best agreement
is obtained with \KandY\ and \RacoonWW; however, 
the measurements are of insufficient statistical precision to 
distinguish between the different calculations. The OPAL result
is two standard deviations below the prediction of \KoralW. Although the
statistical significance is low, the 
${\cal{O}}(\alpha)$ NL electroweak corrections of YFSWW implemented 
in \KandY\ improve the agreement between data and Monte Carlo 
(the dominant effect is the inclusion of radiation from the W-bosons, 
specifically its interference with ISR). 
\begin{table}[htbp]
\renewcommand{\arraystretch}{1}
\begin{center}
\begin{tabular}{|l|c|} \hline 
                               & {Data/Theory} \\ \hline
    \KandY\                   & $0.99 \pm 0.09 \pm 0.04$    \\
    \RacoonWW\                & $0.98 \pm 0.09 \pm 0.06$    \\
    \eewwgammamc\             & $0.91 \pm 0.09 \pm 0.04$    \\
    \KoralW\                  & $0.84 \pm 0.08 \pm 0.04$    \\ \hline
\end{tabular}
\end{center}
\caption{The ratios of the experimental to expected SM $\WWg$ 
         cross-sections averaged over $\roots$ for four theoretical
         calculations. The errors are from the
         statistical and systematic uncertainties  
         on the measurement of the $\WWg$ cross-section.
\label{tab:ratio} }
\end{table}

\begin{figure}[htbp]
 \epsfxsize=14cm 
 \begin{center}
 \epsffile{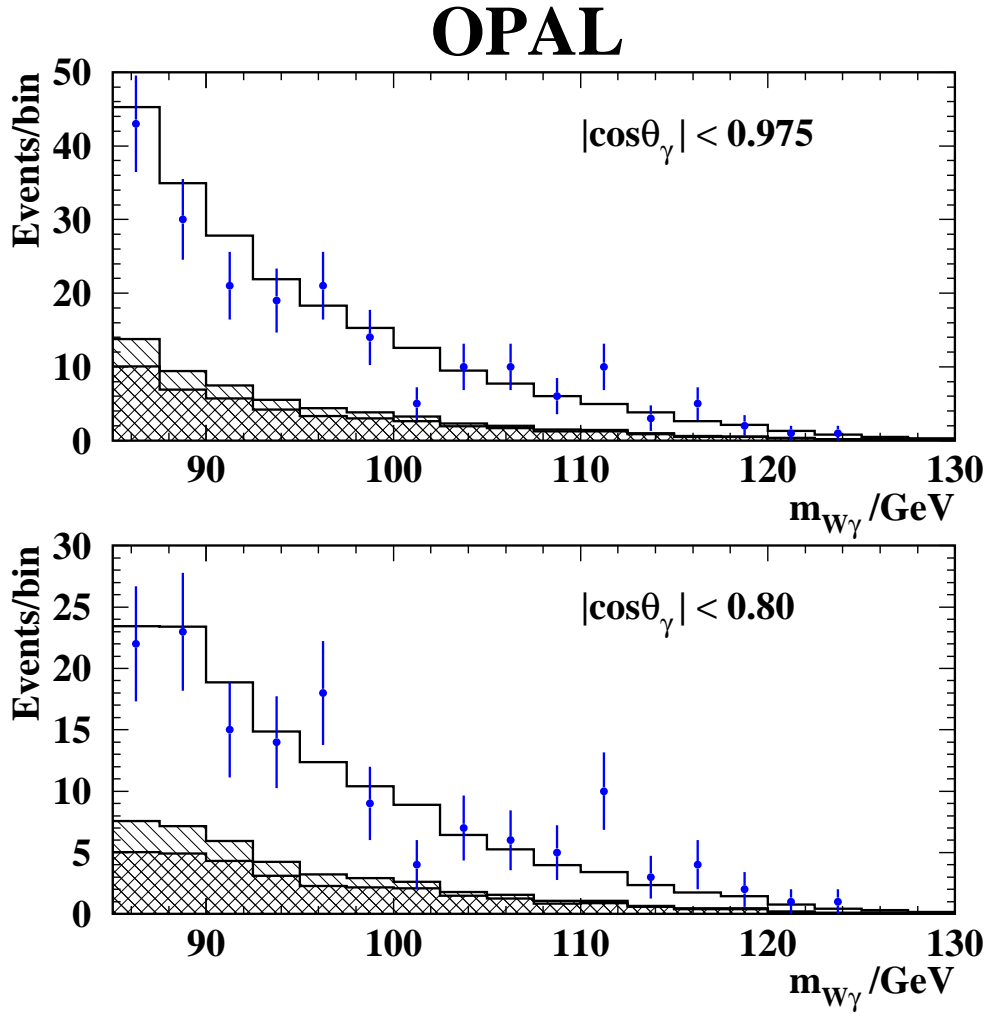}
 \caption{Reconstructed invariant mass of $\mathrm{W}^\pm\gamma$ 
          in selected \WWg\ events with $\Egam>2.5$~GeV 
          (two entries per event). The data are shown by the 
          points, the Standard
          Model expectation, determined from \KandY,  is
          shown by the histogram. The singly hatched histograms 
          show the contribution from FSR from leptons and the doubly hatched
 	  histograms show the background.}
 \label{fig:wstar}
 \end{center}
\end{figure}

New physics could appear as resonant structure in the  
$\mathrm{W}\gamma$ invariant mass distribution (for example
the decay of an excited W-boson, 
$\mathrm{W}^*\rightarrow \mathrm{W}\gamma$). To investigate
this possibility, for $\qqln\gamma$ and $\qqqq\gamma$ candidates
the invariant masses of the two $\mathrm{W}^\pm\gamma$ combinations
in selected \WWg\ events are obtained from an additional 
kinematic fit. The fit uses the constraints of energy and 
momentum conservation and the constraint that the invariant
masses of the reconstructed 
$\fafb$ and $\fcfd$ systems are both equal to the W-mass 
(previously the requirement was that both masses be equal).
Only events for which the kinematic fit converges are retained. For
MC events this cut rejects approximately 16~\% of selected signal events. 
The $\mathrm{W}\gamma$ invariant mass is calculated
from the four-momenta of the four fermions and the
photon returned by the fit.  
Figure~\ref{fig:wstar} shows the reconstructed invariant mass 
distribution for the two $\mathrm{W}^\pm\gamma$ combinations for  
selected \WWg\ events with $\Egam>2.5$~GeV. No resonant structure 
is observed. The 
data from the region $|\cosg|<0.80$, where any contribution 
from new physics might be expected to be most apparent
are also shown.  

\subsection{Systematic Uncertainties}

\label{sec:sys}

The contributions to the systematic uncertainties on
the $\WWg$ cross-sections for the five values of $\roots$ are listed in 
Table~\ref{tab:syserr} and are described below.  The total systematic
errors
are taken as the sum in quadrature of these components. When determining
the average ratio of data to MC the systematic error components for
the five energies are taken to be 100~\% correlated. 

\begin{table}[htbp]
\renewcommand{\arraystretch}{1}
\begin{center}
\begin{tabular}{|l|r|r|r|r|r|r|} \hline 
   \multicolumn{7}{|c|}{Systematic Uncertainty on $\sigWWg$/fb} \\ \hline
   Error Source & Variation &\multicolumn{5}{c|}{ $<\roots>$/GeV} \\ 
                &           & 183 & 189 & 195 & 201 & 206       \\ \hline
   Photons from jets & $\pm15~\%$        & 9 & 10 & 12 & 13 & 13 \\  
   Photon energy scale &$\pm4~\%$        & 6 & 8 & 5 & 9 & 10\\
   Photon angular acceptance &$\pm5$~mrad& 4 & 6 & 4 & 7 & 7 \\ 
   Photon energy resolution &$\pm10~\%$  & 3 & 4 & 3 & 5 & 5 \\
   $\WW$ Selection &$\pm1.1~\%$          &\hspace{3mm} 3 &\hspace{3mm} 4 &\hspace{3mm} 3 &\hspace{3mm} 5 &\hspace{3mm} 5 \\ 
   Photon Identification &$\pm1.0~\%$    & 3 & 4 & 3 & 5 & 5 \\
   Photon Isolation &$\pm1.0~\%$         & 3 & 4 & 3 & 5 & 5 \\
   \qq\ Background & $\pm6.5~\%$         & 2 & 2 & 3 & 3 & 3 \\
   Kinematic Fits &$\pm0.5~\%$           & 1 & 2 & 1 & 2 & 2 \\
   Monte Carlo Statistics &$\pm0.4~\%$   & 1 & 2 & 1 & 2 & 2 \\
   Luminosity & $\pm0.3~\%$              & 1 & 1 & 1 & 1 & 1 \\ \hline
   Total Systematic Error &              & 13  & 17 & 15 & 20  & 21 \\ \hline
   Statistical Error      &              & 117 & 74 & 84 & 100 & 73 \\ \hline
\end{tabular}
\end{center}
\caption{The contributions to the experimental error on the
       $\WWg$ cross-section for the five different values of
       $\roots$. The systematic variations on the various
       sources of error are indicated.\label{tab:syserr} }
\end{table}

\bigskip
\noindent
\underline{\bf\boldmath Modelling of photons from jets}:
The modelling of photon candidates associated with the 
hadronic jets (both from
FSR and from $\pi^0$ and $\eta$ decays)  is studied by 
comparing the rate at which photons are identified in
$\Zzero\rightarrow\qq$ events to the \Pythia\ prediction (for this comparison
data recorded at $\roots\sim\Mz$ during the $1998-2000$ operation of the LEP 
accelerator are used).
For 2.5~GeV$<\Egam<20$~GeV, there are $(38\pm2)~\%$ more photon candidates
identified in the data than expected from the Monte Carlo. 
Above 20~GeV the data are consistent with the Monte Carlo expectation.
The ratio of data to Monte Carlo is used to estimate an 
energy-dependent correction (in photon energy bins of 2.5 GeV) 
to the Monte Carlo expectation for the 
background from \WW\ events with photons associated with jets. 
After the $\WWg$ event selection,  this corresponds to a $(30\pm2)~\%$ 
correction to the background from photons from jets\footnote{For the 
comparison with \RacoonWW\ given in Table~\ref{tab:ratio} the systematic 
errors from photons from jets are calculated differently. In  
\RacoonWW\ it is impossible to separate photons from FSR from quarks from 
other diagrams. Consequently the signal definition is modified to 
include all FSR photons within the theoretical acceptance cuts. 
In this case the data/MC discrepancy for photons from jets in 
$\Zzero\rightarrow\qq$ may either be assigned to a mis-modelling of
FSR (signal) or to a mis-modelling of hadron production rate (background).
Consequently the systematic uncertainties are larger than for the case
when FSR from quarks is also treated as background. 
The central value for the \RacoonWW\ comparison uses the average of the 
results obtained and half the difference is assigned as a systematic error.}.  
Half the size of the correction is propagated as a systematic 
uncertainty. In the evaluation of the other systematic uncertainties 
all comparisons between data and MC are performed after making 
this correction.

\bigskip
\noindent
\underline{\bf\boldmath ECAL energy scale}:
A bias in the energy scale for photons (data relative to Monte Carlo)
in the region of the energy cut, {\em i.e.} $\Egam\sim2.5$~GeV, would result
in a systematic bias in the $\WWg$ cross-section measurement.
The uncertainty on the ECAL energy scale for photons in this region is
estimated by examining photons from $\pi^0$ decays in 
$\epem\rightarrow\qq$ events recorded at $\roots\sim\Mz$ during 1998$-$2000
and  $\epem\rightarrow\qq(\gamma)$ events recorded at $\roots>180$~GeV.
The mean reconstructed $\pi^0$ mass for $\pi^0$ candidates containing
a photon with $2~\GeV<\Egam<3$~GeV is $(142\pm2)$~MeV/$c^2$ in data
compared to 137~MeV/$c^2$ in Monte Carlo. As a result a 4~\% systematic
uncertainty on the ECAL energy scale in the region of $\Egam\sim2.5$~GeV
is assigned. The resulting systematic uncertainty on the 
cross-section is 2~\%.

\bigskip
\noindent
\underline{\bf\boldmath Photon Angular Acceptance}:
The systematic error associated with the requirement
of $|\cosg<0.975|$ depends on the accuracy of the Monte Carlo simulation 
of the angular reconstruction from ECAL clusters at the edge of the 
acceptance. By comparing the reconstructed polar angle from 
different detectors (ECAL, tracking, muon chambers) the ECAL
acceptance is known to $\pm3$~mrad out to $|\cosg|<0.96$. Beyond
the tracking acceptance it is not possible to make this comparison.
Therefore a 5~mrad uncertainty on the edge of the acceptance
is assigned. 

As a cross-check a sample of ISR photons from 
$\epem\rightarrow\qq(\gamma)$ events is used.
Multi-hadronic events recorded at $180~\GeV<\roots<209$~GeV
are selected~\cite{bib:LEP2MH}. Photons are identified using the 
same criteria as for the $\WWg$ cross-section analysis. 
In the data 241 photons are reconstructed in the
region $0.950<|\cosg|<0.975$ compared to the Monte Carlo expectation
of 237.1. A 5~mrad bias between data and Monte Carlo would result
in an expected discrepancy of 28.5 events in this region. The good
agreement between data and Monte Carlo provides confirmation that
the assigned uncertainty of 5~mrad is reasonable.

\bigskip
\noindent
\underline{\bf\boldmath ECAL energy resolution}:
The systematic error from the uncertainty in the 
ECAL energy resolution is obtained in a similar manner as that used for
the ECAL energy scale using the same $\pi^0$ sample. There is no evidence
for a difference between data and Monte Carlo within the statistical
precision of the comparison ($\pm10~\%$). The precision of this comparison
is used to assign a ($10~\%$) uncertainty the energy resolution,
which, when propagated to the uncertainty on the $\WWg$ cross-section
yields a systematic error of $\pm1~\%$.

\bigskip
\noindent
\underline{\bf\boldmath \WW\ selection efficiency}:
Systematic uncertainties in the $\WW$ event selection will result
in corresponding uncertainties in the $\WWg$ event selection. 
The estimated systematic uncertainty on the $\WW$ selection efficiency is
1.1~\%\cite{bib:xs189}, where the largest uncertainties are related to the 
QCD and fragmentation modelling of jets.
For the data sample considered here, the $\WW$ event selection yields 
11752 events which is statistically compatible with the Monte Carlo 
expectation of $11670\pm58$ (where the error is taken to be the theoretical 
uncertainty on the CC03 cross-section). The difference is consistent
with the quoted systematic error of 1.1~\%.

\bigskip
\noindent
\underline{\bf\boldmath Photon Identification}:
A systematic uncertainty of 1~\% is assigned to cover the
uncertainties in the simulation of the photon conversion rate 
and the accuracy of the simulation of the electromagnetic 
cluster shape\cite{bib:invHiggs}. Systematic uncertainties
arising from the isolation requirements are discussed below.
The efficiency obtained from \KandY\ is consistent with that 
from \KoralW\ and no additional systematic uncertainty is
assigned.

\bigskip
\noindent
\underline{\bf\boldmath Photon Isolation}:
The systematic error associated with the isolation requirements
depends on the accuracy of the Monte Carlo simulation of the fragmentation
process in hadronic jets. This is verified in $\Zzero\rightarrow\qq$ events
recorded at $\roots\sim\Mz$ during $1998-2000$. For each selected
event, the inefficiency of the isolation requirements 
is determined for random orientations of the isolation cone and 
parametrised as a function of the angle between the
cone and the nearest jet. For all jet-cone angles the inefficiency
in the Monte Carlo and data agree to better than 1~\%, consequently
a 1~\% systematic error is assigned. Consistent results, albeit with lower
statistical precision, are obtained from $\WW\rightarrow\qqln$ events. 

As a cross-check of the photon identification and isolation
requirements the sample of reconstructed photons in 
$\epem\rightarrow\qq(\gamma)$ events is used. 
The ratio of 
the number of reconstructed photons with $2.5~\GeV < \Egam< 50$~GeV in the
data to the Monte Carlo expectation is $1.015\pm0.023$. 
Good agreement is observed over all $\cosg$. Due to the limited
statistical sensitivity of this test no additional systematic uncertainty 
is assigned to the photon identification/isolation efficiency.

\bigskip
\noindent
\underline{\bf\boldmath $\qq\gamma$ background}:
The dominant source of non-\WW\ background is from  
$\epem\rightarrow\Zgamma\rightarrow\qq\gamma$ where the identified photon
candidate is a genuine photon from ISR. Uncertainties in
the modelling of QCD/fragmentation lead to systematic uncertainties
in the level of background from  $\epem\rightarrow\qq\gamma$ events in the
\WW\ event selection\cite{bib:xs189}. As a result the $\qq\gamma$ 
background in the \WWg\ selection is uncertain to 6.0~\%. An additional
systematic error of 2.5~\% arises from the uncertainties in the
modelling of ISR in  $\epem\rightarrow\qq\gamma$ events.    

\bigskip
\noindent
\underline{\bf\boldmath Kinematic Fits}:
The \WWg\ event selections require that a kinematic fit converges 
and has a reasonable probability. Possible  mis-modelling of the 
detector response/resolution could result in a difference in the rates
at which the fits fail for data and Monte Carlo. This was checked by 
applying the kinematic fits used in the W mass analysis to all selected
\WW\ events and comparing the failure rates for data and Monte Carlo.
The efficiency (\qqlv\ and \qqqq\ combined) in data is $(83.9\pm0.4)~\%$
compared with the Monte Carlo expectation of 84.1~\%. The ratio
of these efficiencies is $0.997\pm0.005$ and, consequently, a systematic
uncertainty of 0.5~\% is assigned.

\bigskip
\noindent
\underline{\bf\boldmath Monte Carlo Statistics}:
The effect of finite Monte Carlo statistics is taken into account and
leads to 0.3~\% systematic uncertainties on the measured cross-sections.

\bigskip
\noindent
\underline{\bf\boldmath Luminosity}:
The total uncertainty on the integrated luminosity of the data samples
is 0.3~\%, dominated by systematics.

\section{\boldmath Limits on $\Mw$ $\cal{O}(\alpha)$ Systematic Uncertainties}

The anticipated experimental error on $\Mw$ from LEP2 is 
approximately 35~MeV. A potential source of theoretical uncertainty
is the treatment of higher order QED corrections in the Monte Carlo
programs used to simulate the process $\epem\rightarrow4f(\gamma)$.
A recent estimate suggests a total theoretical systematic uncertainty
due to $\cal{O}(\alpha)$ effects of 5~MeV\cite{bib:sysMw}. 
However, as pointed out by the authors\cite{bib:sysMw}, 
this estimate is based upon the invariant mass of the 
$\mu^-{\overline{\nu}}_\mu$ system in 
$\epem\rightarrow\mu^-{\overline{\nu}}_\mu u{\overline{d}}(\gamma)$ events,
whereas the experimental procedure used to extract $\Mw$ is  
complicated by the fact that the four LEP collaborations
use kinematic fits to improve significantly the event-by-event 
W-mass resolution\cite{bib:MwA,bib:MwD,bib:MwL,bib:MwO}. 
One effect of the kinematic fit is to constrain the total energy of the 
reconstructed fermions to $\roots$. For events with photons from ISR 
this procedure introduces a bias in the reconstructed W-mass as the 
energies of four fermions should be constrained to $\rootsprime$, 
the centre-of-mass energy after photon radiation, rather than to $\roots$.
Consequently, as a result of the experimental procedure 
used to extract
$\Mw$, the $\cal{O}(\alpha)$ theoretical systematic uncertainties 
may be significantly greater than those obtained by considering the 
invariant mass distribution of the final state fermions\cite{bib:Thomson}. 

\subsection{\boldmath QED and Electroweak Corrections in KandY}

In the \KandY\ generator it is possible to study the effects of different
theoretical corrections using event correction weights\cite{bib:KandY} which, 
when used to weight generated events, allow different theoretical
predictions to be tested. By processing generated
fully-simulated events through the full 
\Opal\ W-mass analysis it is possible to determine the W-mass biases 
associated with these corrections. For example, degrading the 
${\cal{O}}(\alpha^3)$ exponentiated LL treatment of collinear ISR to 
${\cal{O}}(\alpha^2)$ results in a systematic bias of less than 
1~MeV\cite{bib:sysMw,bib:MwO}. In a similar manner the non-leading
(NL) ${\cal{O}}(\alpha)$ electroweak corrections, including radiation 
from the W-bosons, may be switched off using the appropriate event 
correction weights, $w^i_{{\overline{NL}}}$. When applied to the full 
\Opal\ W-mass analysis it is found that dropping the ${\cal{O}}(\alpha)$ NL 
electroweak corrections results in a shift in the reconstructed W-mass of 
15~MeV. This relatively large bias is due to the modification of the 
$\rootsprime$ distribution rather than a distortion in the invariant mass 
distribution of the fermion pairs\cite{bib:Thomson}. The change in the 
$\rootsprime$ distribution is due to the inclusion in \KandY\ of 
the diagrams for
radiation from the W-bosons which, through interference with the  
ISR diagrams, reduces the cross-section for the production of real 
photons\cite{bib:Plazek}\footnote{This effect
was investigated by running YFSWW with {\tt KEYCOR=2} and {\tt KEYCOR=3}
switching between the YFS form factor solely for ISR and the full
form factor including WSR and interference between
WSR and ISR.}. Although the fractional change in $\WWg$ cross-section 
is largest at $\cosg=0$ where the photon production rate is 
reduced by 30~\%\cite{bib:KandY}, in absolute terms the reduction
in the cross-section shows no strong $\cosg$ dependence.
Consequently the $\WWg$ cross-section measurement provides a test
of the modelling of radiation from the W-bosons 
(and the interference with ISR) in the KandY Monte Carlo.
Since the largest source of systematic
bias from the so-called ${\cal{O}}(\alpha)$ NL corrections is a direct 
result of the modification of the spectrum of real photons, the
associated systematic uncertainties on $\Mw$ 
may be constrained by the measurement of the $\WWg$ cross-section.

\subsection{\boldmath Constraints from the $\WWg$ Measurements}  

To investigate the experimental limits on possible biases on the 
measurement of $\Mw$ due to photon production away from the collinear
region the correction weights from \KandY\ are first expressed in
the form
\begin{eqnarray*}
    w^i_{\overline{NL}} & = & 1.0 + \delta^i_{\overline{NL}}.
\end{eqnarray*}
By modifying the weights to
\begin{eqnarray*}
    w^i_{\overline{NL}} & = & 1.0 + \kappa \delta^i_{\overline{NL}}
\end{eqnarray*}
it is possible to investigate a continuous range of scenarios.
A value of $\kappa=0$ corresponds to the treatment of ${\cal{O}}(\alpha)$ NL electroweak
corrections of YFSWW ({\em i.e.} the default in \KandY) and $\kappa=1.0$
corresponds to dropping the NL ${\cal{O}}(\alpha)$ 
corrections.  
The parameter $\kappa$ and its errors 
are obtained from a binned extended maximum likelihood fit to the
$|\cosg|$ distribution of Figure~\ref{fig:result_eg}, taking into account
both the overall normalisation and shape, giving  
\begin{eqnarray*}
  \kappa & = & 0.38 \pm 0.45 \pm 0.15,
\end{eqnarray*}
where the first error is statistical and the second due to systematic
uncertainties in the event selection efficiency. 
The data favour the \KandY\ prediction including the NL corrections.
Most of the sensitivity comes from the photon rate rather than the
angular distribution.
The measured value of $\kappa$ suggests that the measured 
value of $\Mw$ from the \Opal\ W-mass analysis, obtained using \KandY\ as a 
reference, should be corrected by $(-5\pm6)$~MeV. 
Using the measured cross-section alone
gives a similar result of $(-1\pm7)$~MeV.
From these studies it is concluded that the systematic error on $\Mw$ due
to the Monte Carlo implementation of QED diagrams resulting in real photon
production away from the collinear region should be not more than $6$~MeV.

\section{Anomalous Quartic Gauge Boson Couplings}

\label{sec:qgc}

The non-Abelian nature of the electroweak sector of the Standard Model  
results in vector boson self-interactions. In addition to the
triple gauge boson couplings (TGCs), $\WW\gamma$ and $\WW\Zzero$, 
the Standard Model predicts the existence of four quartic gauge couplings, 
$\WW\WW$, $\WW\ZZ$, $\WW\Zzero\gamma$ and $\WW\gamma\gamma$. 
These couplings are not expected to play a significant role at LEP
energies, but will be important at the LHC~\cite{bib:LHC} 
and at a future TeV linear collider~\cite{bib:Boos}.
 
\begin{figure}[ht]
 \begin{center}
 \epsfxsize=7cm
 \epsffile{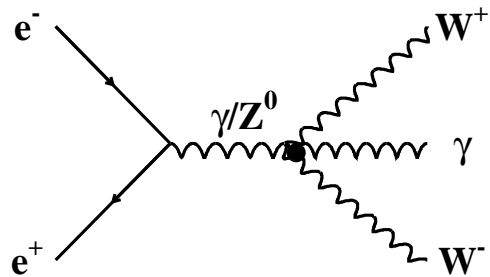}
 \caption{Standard Model production diagram for 
    the $\WW\gamma$ final states involving the $\WW\gamma\gamma$ and 
        $\WW\Zzero\gamma$ quartic gauge couplings.}
 \label{fig:qgcdiag}
 \end{center}
\end{figure}

Quartic gauge boson couplings can be probed in final states with three 
vector bosons. At LEP centre-of-mass energies, final states involving 
three massive gauge bosons are kinematically out of reach. However, it 
is possible to study the processes 
$\epem\rightarrow\WW\gamma$\cite{bib:OPALwwg,bib:L3wwg} and
$\epem\rightarrow\Zzero\gamma\gamma$\cite{bib:L3qqgg}. 
In the Standard Model,
the contribution of the quartic couplings
to $\epem\rightarrow\WW\gamma$, shown in Figure \ref{fig:qgcdiag}, 
is expected to be too small to measure and that to 
$\epem\rightarrow\Zzero\gamma\gamma$ is zero. Nevertheless, it is 
possible to set direct limits on possible anomalous contributions to
the quartic gauge boson couplings. 

\subsection{Theoretical Framework}

In the SM the form and strength of the vector boson 
self-interactions are fixed by SU(2)$\times$U(1) gauge invariance.
As is the case for triple gauge boson couplings~\cite{bib:Hagiwara}, 
in extensions to the SM, anomalous quartic couplings
can be parametrised by additional terms in the 
Lagrangian~\cite{bib:Berlanger,bib:Stirling1,bib:eewwg}.
These are required to conserve 
custodial SU$(2)_c$ symmetry in order to avoid deviations of the 
$\rho$ parameter\footnote{$\rho = \Mws/(\Mzs\cos^2\theta_W)$, where \Mw\ 
and \Mz\ are the masses of the W$^\pm$ and \Zzero\ bosons and 
$\theta_W$ is the weak mixing angle.} 
from the experimentally well established value close to 1. 
Only operators which do not introduce anomalous triple gauge 
couplings are considered. For example, the anomalous quadrupole moment 
operator generates both \WW$\gamma$ and \WW$\gamma\gamma$ couplings. 
Therefore, it is not considered as a source of genuine anomalous quartic 
couplings since its strength, $\lambda_\gamma$, is already tightly 
constrained from the study 
of TGCs at LEP~\cite{bib:tgc189,bib:leptgc} and at 
the Tevatron~\cite{bib:tevatgc}. The lowest dimension operators 
which generate genuine anomalous quartic couplings involving photons are 
of dimension six. Three such possibilities are considered here, 
${\cal{L}}_6^0$, 
${\cal{L}}_6^c$~\cite{bib:Berlanger} and 
${\cal{L}}_6^n$~\cite{bib:Eboli,bib:eewwg}:
\begin{eqnarray*}
     {\cal{L}}_6^0 & = & 
             -\frac{e^2}{16\Lambda^2}a_0 \Fmv \FMV \Wa.\WA, \\
     {\cal{L}}_6^c & = & 
             -\frac{e^2}{16\Lambda^2}a_c \Fma \FMB \Wb.\WA, \\
     {\cal{L}}_6^n & = &       
              i\frac{e^2}{16\Lambda^2}a_n \epsilon_{ijk} \Wma^{(i)} 
               \WV^{(j)}\Wka\Fmv, 
\end{eqnarray*}
with 
\begin{eqnarray*}
     \WM & = & \left(  \begin{array}{c}
                         \frac{1}{\sqrt{2}}( W_\mu^+ + W_\mu^-) \\ 
                         \frac{i}{\sqrt{2}}( W_\mu^+ - W_\mu^-) \\
                         Z_\mu / \cos\theta_{W} 
                       \end{array}
               \right),
\end{eqnarray*}
where $\Fmv$ and $\Wmv$ are the field strength tensors of the
photon and $W$ fields respectively. 
Both ${\cal{L}}_6^0$ and ${\cal{L}}_6^c$, which conserve 
$C$ and $P$ (separately), generate 
anomalous $\WW\gamma\gamma$ and $\ZZ\gamma\gamma$ couplings. 
The $CP$ violating term
${\cal{L}}_6^n$ results in an anomalous $\WW\Zzero\gamma$ 
coupling.
In each case, the strength of the
coupling is proportional to $a_i/ \Lambda^2$, where $\Lambda$ 
represents a scale for new physics. 
A more general description of the operators leading to 
anomalous quartic couplings accessible at LEP can be found 
in the paper of B\'elanger \etal\cite{bib:lep2boudjema}. 
The two additional dimension 6 operators, parametrised by 
$\hat{a}_0$ and $\hat{a}_c$, identified by 
Denner~\etal\cite{bib:RacoonQGC} are not considered here as 
the  effects of $\hat{a}_0$ and $\hat{a}_c$
are almost identical to those of ${a}_0$ and ${a}_c$, respectively.

\subsection{Experimental Limits}

The selected $\WW\gamma$ events are used to set limits on possible
anomalous contributions to the  $\WW\gamma\gamma$ and 
$\WW\Zzero\gamma$ quartic gauge couplings. The limits are extracted 
from the measured differential cross-section as a function of 
the photon energy and photon polar angle.
The signal of anomalous quartic gauge boson couplings at LEP would
be an excess of \WWg\ events. The effect of anomalous QGCs increases  with 
photon energy. Furthermore, the sensitivity to anomalous QGCs increases 
with increasing $\sqrt{s}$.

\begin{figure}[htbp]
 \epsfxsize=\textwidth
 \epsffile{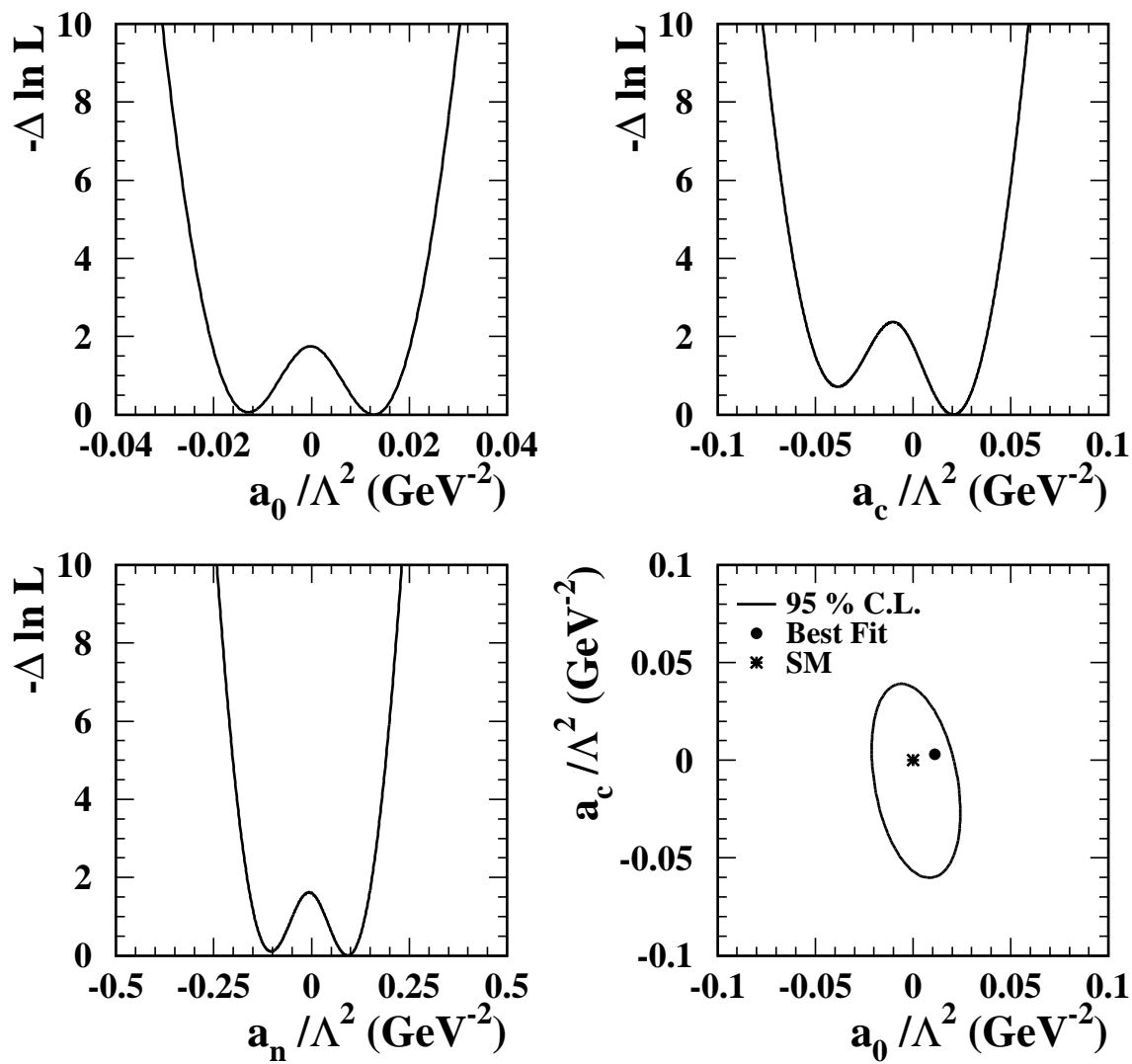}
 \caption{Likelihood curves for the anomalous QGC parameters
  $a_0$, $a_c$ and $a_n$. Also shown is the 95~\% C.L. region 
  for  ($a_0$,$a_c$). The curves include the experimental systematic
  uncertainties and a 10~\% theoretical uncertainty for the
   $\epem\rightarrow\WWg$ cross-section.}
 \label{fig:qgclimits}
\end{figure}

The calculation of Stirling and Werthenbach~\cite{bib:eewwg} 
allows for the assessment of the impact of anomalous quartic couplings
and is implemented in the \eewwg\ program.
This calculation includes the ISR diagrams, the WSR diagrams, the
SM QGC diagram and can accommodate anomalous quartic couplings.
However, the recent implementation of anomalous QGCs in the 
RacoonWW\cite{bib:RacoonQGC} and WRAP\cite{bib:WRAP} programs 
identified a problem with the \eewwg\ program, indicating 
that  $a_0 \rightarrow -a_0$ and $a_c \rightarrow -a_c$ 
in \eewwg. In this study the \eewwg\ program is used with the
signs of $a_0$ and $a_c$ inverted.
To set limits on possible anomalous couplings a binned maximum 
likelihood fit to the observed distribution of [\Egam, $|\cosg|$] is 
performed using bins of [5~GeV, 0.1]. Fits are performed to the data for the five 
separate energy ranges of Table~\ref{tab:lumi} 
and the resulting likelihood curves are summed. The effects of anomalous 
couplings are introduced by reweighting events generated with 
$\KandY$ using the average ratio of anomalous QGC to SM matrix elements
from \eewwg\ in the relevant bin of [\Egam, $|\cosg|$].
The resulting summed likelihood curves are shown in 
Figure~\ref{fig:qgclimits}. 
Results are obtained for three single parameter fits, where one 
of $a_0$, $a_c$ or $a_n$ is varied whilst the other two parameters
are set to zero, and a two parameter fit to $\{a_0,a_c\}$. 
The results include the effect of the
experimental systematic errors 
and assume a 10~\% theoretical uncertainty\footnote{This represents
a conservative estimate of the theoretical uncertainty, comparisons of
YFSWW and RacoonWW suggest 5~\%.}
on the cross-section for $\epem\rightarrow\WWg$. These uncertainties
are taken to be 100~\% correlated between the five energy ranges. 
The best fit does not occur
at the SM value of zero. However this does not imply the data are 
inconsistent with the SM. The consistency with the SM prediction, 
given by the probability of obtaining a value of $-\ln{\cal{L}}$ 
greater than that observed for $\{a_0=0,a_c=0,a_n=0 \}$, is 19~\%.  
The 95~\% confidence level upper limits on the anomalous couplings,
obtained from the likelihood curves, $\Delta(\ln{\cal{L}})=1.92$, are:
\begin{eqnarray*}
   -0.020~\mathrm{GeV}^{-2} < & a_0/ \Lambda^2 & < 0.020~\mathrm{GeV}^{-2}, \\ 
   -0.053~\mathrm{GeV}^{-2} < & a_c/ \Lambda^2 & < 0.037~\mathrm{GeV}^{-2}, \\
   -0.16~\mathrm{GeV}^{-2} < & a_n/ \Lambda^2 & < 0.15~\mathrm{GeV}^{-2},
\end{eqnarray*}
For $a_c$ the region 
$-0.020~\mathrm{GeV}^{-2} < a_c/ \Lambda^2  < -0.002~\mathrm{GeV}^{-2}$ is 
also excluded at the 95~\% C.L.
The derived upper limits are less restrictive than the expected limits. 
For example, the expected limit on $a_0$ is 
$|a_0/ \Lambda^2| < 0.014~\mathrm{GeV}^{-2}$. 
The limits are worse than expected due to 
a slight excess of high energy photons in the $\roots>205$~GeV data sample.

\section{Conclusions}

Using  \Nobs\ \WWg\ candidates with photon energies greater than 2.5~GeV
the $\WW\gamma$ cross-section is measured at five values of $\roots$.
The results are consistent with the Standard Model expectation. 
Averaging over the five energies, the ratio of the observed cross-section
to the prediction of the concurrent Monte Carlo KoralW and YFSWW
(KandY) is
\begin{eqnarray*}
    R({\mathrm{data/MC}})    & = & 0.99 \pm 0.09 \pm 0.04,
\end{eqnarray*}
where the errors represent the statistical and systematic uncertainties
respectively. This provides a 10~\% test of the KandY implementation
of ${\cal{O}}(\alpha)$ effects producing a real photon away from
the collinear region. 
From these studies it is concluded that the systematic error on $\Mw$ due
to the Monte Carlo implementation of QED diagrams resulting in real photon
production away from the collinear region should be not more than $6$~MeV.

The data are used to derive 95~\% confidence level upper limits 
on possible anomalous contributions to the
$\WW\gamma\gamma$ and $\WW\Zzero\gamma$ vertices:
\begin{eqnarray*}
   -0.020~\mathrm{GeV}^{-2} < & a_0/ \Lambda^2 & < 0.020~\mathrm{GeV}^{-2}, \\ 
   -0.053~\mathrm{GeV}^{-2} < & a_c/ \Lambda^2 & < 0.037~\mathrm{GeV}^{-2}, \\
   -0.16~\mathrm{GeV}^{-2} < & a_n/ \Lambda^2 & < 0.15~\mathrm{GeV}^{-2},
\end{eqnarray*}
where $\Lambda$ represents the energy scale for new physics.

\section{Acknowledgments}

We would like to thank James Stirling and Anja Werthenbach for providing 
the program \eewwg\ which is used to determine the effects of anomalous 
couplings in \WWg\ events. We also greatly appreciate their many 
useful suggestions and comments. We would also like to thank Markus Roth 
and the RacoonWW group for providing cross-section calculations using the 
\RacoonWW\ program. We thank Marius Skrzypek and Wieslaw 
P{\l}aczek for many useful discussions.  

\par
We particularly wish to thank the SL Division for the efficient operation
of the LEP accelerator at all energies
 and for their close cooperation with
our experimental group.  In addition to the support staff at our own
institutions we are pleased to acknowledge the  \\
Department of Energy, USA, \\
National Science Foundation, USA, \\
Particle Physics and Astronomy Research Council, UK, \\
Natural Sciences and Engineering Research Council, Canada, \\
Israel Science Foundation, administered by the Israel
Academy of Science and Humanities, \\
Benoziyo Center for High Energy Physics,\\
Japanese Ministry of Education, Culture, Sports, Science and
Technology (MEXT) and a grant under the MEXT International
Science Research Program,\\
Japanese Society for the Promotion of Science (JSPS),\\
German Israeli Bi-national Science Foundation (GIF), \\
Bundesministerium f\"ur Bildung und Forschung, Germany, \\
National Research Council of Canada, \\
Hungarian Foundation for Scientific Research, OTKA T-038240, 
and T-042864,\\
The NWO/NATO Fund for Scientific Research, the Netherlands.\\

\section{Appendix}

For the purpose of the combination of results from the four 
LEP experiments cross-section results are obtained for the
signal definition:
\begin{itemize}
  \item $\Egam   > 5$~GeV,
  \item $|\cosg| < 0.95$,
  \item $\cosgf  < 0.90$,
  \item $|\Mfafb-\Mw|$ and $|\Mfcfd-\Mw| < 2\,\Gw$.
\end{itemize}
The experimental cuts on the photon acceptance are modified to
match the signal definition. For the modified selection, 
124 events are selected, compared to the SM expectation (KandY)
of $118.7\pm0.6$. The results are summarised in Table~\ref{tab:lepxsec}.
\begin{table}[htbp]
\renewcommand{\arraystretch}{1}
\begin{center}
\begin{tabular}{|c|rcrcl|c|} \hline
                   & \multicolumn{6}{c|}{$\sigWWg$/fb} \\ 
   $<\roots>$/GeV  & \multicolumn{5}{c}{Data} & $\KandY$ \\ \hline
   182.68          & 102 &\hspace{-3mm}$\pm$& \hspace{-4mm} 60& \hspace{-3mm}$\pm$& \hspace{-4mm}5 & $141\pm2$  \\
   188.63          & 163 &\hspace{-3mm}$\pm$& \hspace{-4mm} 41& \hspace{-3mm}$\pm$& \hspace{-4mm}6 & $175\pm3$  \\
   194.44          & 166 &\hspace{-3mm}$\pm$& \hspace{-4mm} 57& \hspace{-3mm}$\pm$& \hspace{-4mm}7 & $201\pm2$  \\
   200.21          & 214 &\hspace{-3mm}$\pm$& \hspace{-4mm} 60& \hspace{-3mm}$\pm$& \hspace{-4mm}7 & $216\pm3$  \\
   205.92          & 298 &\hspace{-3mm}$\pm$& \hspace{-4mm} 50& \hspace{-3mm}$\pm$& \hspace{-4mm}8 & $226\pm3$  \\ \hline
\end{tabular}
\end{center}
\caption{\WWg\ cross-section measurements for the signal definition to be  
         used for a LEP combination of results. The errors are
         statistical and systematic respectively. The systematic uncertainties
         are calculated as described in Section~\ref{sec:sys}.
\label{tab:lepxsec} }
\end{table}

\newpage

\end{document}